\documentclass[12pt]{article}

\usepackage{newtxtext,newtxmath}

\usepackage{ulem}

\usepackage{graphicx}
\usepackage{braket}

\usepackage[letterpaper,margin=1in]{geometry}

\renewenvironment{abstract}
	{\quotation}
	{\endquotation}

\date{}

\makeatletter
\renewcommand{\fnum@figure}{\textbf{Figure \thefigure}}
\renewcommand{\fnum@table}{\textbf{Table \thetable}}
\makeatother

\usepackage{scicite}

\usepackage{mathtools}

\usepackage{url}

\usepackage{hyperref}

\newcommand{\an}[1]{{\color{black}#1}}
\newcommand{\dk}[1]{{\color{black}#1}}

\newcommand{\annew}[1]{{\color{black}#1}}

\def\scititle{
	Spin-force from a Nitrogen-Vacancy ensemble drives a 100 mg levitated resonator}
\title{\bfseries \boldmath \scititle}

\author{
	Anshuman Nayak\textsuperscript{1*},
	Daehee Kim\textsuperscript{1},
	Shilu Tian\textsuperscript{1,2*},
      Jason Twamley\textsuperscript{1}
      \\[0.5em]
	 \small\textsuperscript{1} Quantum Machines Unit, Okinawa Institute of Science and Technology Graduate University, \\ \small Onna, Okinawa 904-0495, Japan\\
\small\textsuperscript{2} Department of Physics, Institute of Science Tokyo, Tokyo 152-8551, Japan\\
\small \textsuperscript{*} Corresponding authors: anshuman.nayak@oist.jp,  shilu.tian@oist.jp}

\begin{document} 

\maketitle

\begin{abstract}  \bfseries \boldmath
The force experienced by a spin in a magnetic field gradient underlies many proposals for hybrid quantum systems. These include schemes for mechanically mediated quantum gates, spin squeezing, searches for exotic forces, and motional superpositions for probing the interface between quantum and gravity. Yet, experimentally observing this spin-force for anything larger than atomic scales has proved challenging. In our work, we demonstrate controllable Center-of-Mass motion of a 
$128 \rm\: mg$
diamagnetically levitated oscillator due to force from an ensemble of Nitrogen-Vacancy (NV) defects in diamond. We induce coherent motion in the oscillator by periodic optical initialisation of the NV spin states, achieving motional amplitudes exceeding 
$100 \rm\:nm$
. Our results mark a key milestone towards spin-based engineering of motional states deep in the high-mass regime.\\[1em]
\end{abstract}

\section{Introduction}
The coupling of spin to mechanical motion has been a central element in the development of quantum mechanics. Starting with the initial experiments of Otto Stern and Walther Gerlach in 1921-22 \cite{Gerlach1922DerMagnetfeld}, researchers initially used the spin-mechanical coupling provided by a spin moving in a magnetic gradient to probe the magnetic moment of iron atoms \cite{Klabunde1934TheIron}, the proton \cite{Estermann1937TheProton}, the electron \cite{Koppe1948DasElektrons}, polarized neutrons \cite{Sherwood1954Stern-GerlachNeutrons}, and various elemental materials \cite{Bucher1991MagneticBeam, Douglass1993MagneticClusters}, which led to the development of matter wave interferometers \cite{Robert1991AtomicAtoms}. Rabi considered the spin-mechanical coupling in the Stern-Gerlach (SG) experiment to be ``astounding'' \cite{Rabi1988OttoQuantization}, and more recently, researchers have developed atomic chips to demonstrate SG manipulation of neutral atoms \cite{MacHluf2013CoherentChip, Wu2019SternGerlachLattice}, and matter wave interferometers \cite{Amit2019T3Interferometer}. 
Spin-mechanical coupling via magnetic \cite{Rabl2009StrongResonator, Abdi2017hBN},  strain \cite{Teissier2014StrainOscillator, Ovartchaiyapong2014DynamicResonator}, or electric fields \cite{Fedele2024CouplingMotion} has led to many investigations of ``hybrid'' quantum systems \cite{Wallquist2009HybridEngineering}, coupling  spins to mechanical cantilevers \cite{Mamin2004SingleMicroscopy, ShimonKolkowitz2012CoherentQubit}, nanomechanical \cite{Hong2012CoherentSpin, Chotorlishvili2013EntanglementResonator},  or electromechanical systems \cite{Li2015HybridCavities, Fedele2024CouplingMotion}, for use in metrology \cite{Xia2016,Zhang2021QuantumNanodiamond, Zhao2022InertialDiamond} and quantum information processing \cite{Fung2024TowardTransport,Arrazola2024TowardPhonons}, and quantum interconnects \cite{Chia2024HybridState}. {\color{black} Many of these ``hybrid'' schemes  use the spin as a sensor of some mechanical motion, most prominent being strain \cite{Teissier2014StrainOscillator, Ovartchaiyapong2014DynamicResonator}, but there is a negligible spin-dependent force on the mechanics.}
{\color{black} Over the past decade, researchers have proposed using a Stern-Gerlach-type (SG) force to couple the center-of-mass (COM) motion of a levitated particle to spins embedded within it \cite{Yin2013, Scala2013, Bose2017, Zhou2025Spin-dependentSuperpositions}, to engineer macroscopic spatial superpositions \cite{Yin2013, RamanNair2025MassiveMagnetomechanics}.
Levitated systems in vacuum are an ideal platform to achieve this due to ultralow mechanical damping and a high degree of control \cite{Gonzalez-Ballestero2021Levitodynamics:Vacuum}.
Macroscopic motional superpositions have the potential to perform ``tabletop'' tests on the quantum nature of gravity   \cite{Bose2017, Marletto2017, Bose2025MassiveGravity},  to test the boundary of quantum mechanics \cite{Qin2019ProposalResonators, Sahoo2023EmergenceSelfGravity, Grossardt2024Self-gravitationalTrajectories}, search for exotic forces or dark matter \cite{Carney2020ProposalMatter, Monteiro2020a, Figueroa2021DarkNetworks}, as well as practical applications in quantum transduction \cite{Fung2024TowardTransport}, and sensing. Most of these theoretical proposals use a single spin embedded within a levitated nanoparticle, but one can use a spin ensemble to generate the spatial superposition as long as the coherence time of the spin ensemble is sufficiently long compared to the mechanical frequency. Indeed, the experimentally demonstrated strong coupling of magnons in a sphere of Yttrium-Iron-Garnett (YIG), to microwaves in cavity-magnonics \cite{Zhang2014}, is established through the collective excitation of a huge spin ensemble $\sim 10^{16}$, in the YIG, which remains coherent for microseconds. 

Nitrogen vacancy defects in diamond have emerged as an excellent system to engineer spin-mechanical couplings.
Recent experimental advances have shown spin manipulation and readout of NV spins in levitated nanodiamonds, albeit without spin-mechanical coupling  \cite{Neukirch2015Multi-dimensionalNanodiamond, Hoang2016ElectronVacuum}, and more recently, the generation of spin-dependent torque and rotational cooling in microdiamonds held in a Paul-trap with embedded NV spins 
\cite{Delord2020, Pellet-Mary2021MagneticInteractions, Perdriat2021Spin-mechanicsParticles, Perdriat2025Spin-dependentMicrolever}.  
However, the detection of the spin-dependent translational motion, an ingredient essential for many of the above-mentioned proposed applications, using any form of mechanical oscillator, levitated or otherwise, has remained elusive, to the best of our knowledge.
}

Here, we report a direct observation and manipulation of \dk{the} COM motion of a 100 milligram-scale diamagnetically levitated oscillator driven \an{purely} by {\color{black} the spin-dependent force of an NV ensemble.}  A diamond slab containing NV spins is rigidly suspended from a levitating graphite plate into a magnetic field gradient generated by a permanent magnet. 
We spin-polarise the NV spin ensemble by green laser illumination and then allow it \dk{to} thermally relax, periodically at the mechanical frequency, leading the spin magnetic moment of the ensemble to switch between optically and thermally polarised states, which in the presence of a \dk{magnetic field} gradient gives rise to a periodic spin-force that then drives the center of mass of the oscillator. 
We interferometrically observe 
this COM motion of the 
oscillator, which can oscillate as large as $\sim 100$ nm in air \an{(and up to $\sim1.5$ $ \mu$m in vacuum)} due to this spin-force. 
\an{We model the force arising from the optical initialisation of NV centers using a seven-level scheme for NV polarisation, which captures the dependence of the force on laser power as well as the magnet diamond separation.}

\section{Results}
\subsection{Setup and operation of the experiment}

\begin{figure} 
	\centering
	\includegraphics[width=\textwidth]{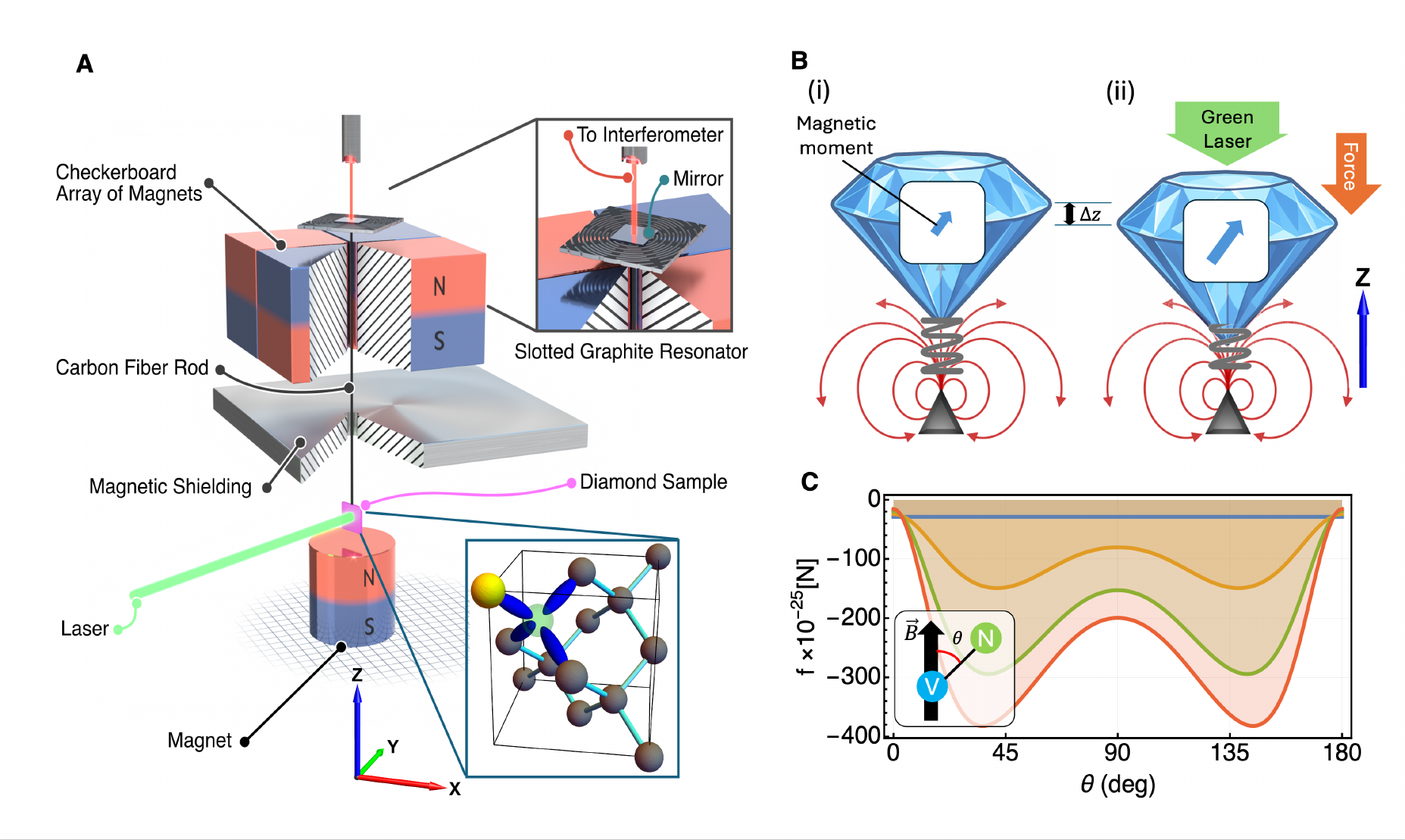}

	\caption{\textbf{\dk{Spin-mediated mechanical motion in a diamagnetic levitation–NV center hybrid system
    }:}
    (\textbf{A}) Schematic of the experimental setup. A diamagnetic slotted graphite slab 
    is levitated above four NdFeB magnets in a checkerboard magnetization arrangement. The diamond plate with NV centers is hung below via a carbon fiber rod through a hole through the magnets. 
    A small mirror on the graphite allows the motion to be monitored via a laser interferometer.
    A cylindrical magnet below the diamond generates a gradient magnetic field. When a $532 \:\rm nm$ green laser illuminates the diamond, NV spin polarization and a magnetic moment are created, which shifts the resonator towards the magnet. 
    (\textbf{B}) 
    (i) With the green laser off, the NV centers are in a thermal spin-state with a small magnetic moment, inducing a small displacement of the diamond towards the magnet. (ii) With the laser on, the induced magnetic moment increases dramatically, resulting in a large displacement towards the magnet, $\Delta z$, from the thermal equilibrium position. Pulsing on/off the green laser at the mechanical resonance frequency magnifies the mechanical response. 
    (\textbf{C}) The simulated force $f$ on a single NV in a magnetic field gradient with {\color{black}  $\mathbf{B}= (B_0+b z)\,\hat{z}$, $B_0\sim 0.63$ T, $b\sim -98 \rm \: T/m$, along the $z$-direction as a function of $\theta$ (NV positioned $z=1$ mm above a cylindrical magnet)}     with illumination of 532 nm light at intensities $I=(0, 10, 30, 50)\;{\rm mW/mm^2}${\color{black}, for blue, orange, green, red line, respectively}.  
    The force is directed towards larger $|\textbf{B}|$, i.e.\dk{,} the NV acts as a paramagnet.\\  }

	\label{fig:fig1} 
\end{figure}

We demonstrate spin-mechanical coupling between an ensemble of nitrogen-vacancy (NV) spins in diamond and a diamagnetically levitated macroscopic mechanical oscillator. The system consists of a slotted graphite plate that diamagnetically levitates above a checkerboard array of permanent magnets and is connected rigidly through a magnetic shielding to a millimeter-scale diamond containing NV centers ($\sim 4.5$ ppm) (Fig.\ref{fig:fig1}). The combined structure forms an integrated mechanical resonator with a total mass of $128$ mg, a center-of-mass resonance frequency of $17.6$ Hz along the vertical $z$ direction and a Q of 55. The motion of the resonator is monitored with an optical interferometer that measures the displacement of a mirror mounted on the graphite plate.

A cylindrical permanent magnet placed beneath the diamond generates both a static magnetic field and a magnetic field gradient at the location of the NV ensemble. 
\dk{
In the presence of the magnetic field gradient, the magnetic moment associated with each spin state gives rise to a state-dependent force.}
We irradiate the green laser at the same frequency as the mechanical oscillation to drive the resonator resonantly to enhance the mechanical response. \an{More experimental details can be found in Materials and Methods section.}

\subsection{Theory of light-induced spin force by NV centers in diamond}
To model the spin force, we consider an ensemble of NV centers oriented along the four crystallographic axes, with the laboratory $z$-axis aligned to $[0,0,1]$ (see Fig. \ref{fig:fig1} (A)). Each NV axis is at an angle of  $\sim 54.7^\circ$ with respect to the laboratory $z$-axis frame, \an{where \{$x,y$\}=\{0,\:0\} is the $z$-axis passing through the axis of the cylindrical magnet}. The Hamiltonian of a single NV center in the lab frame is $H_{\text{NV}} = h D S_m^2 + h \gamma_e \mathbf{B}(
\an{x,y,}\:z)\cdot \mathbf{S}$,
where $S_m$ is the NV $S_z$ operator rotated to the lab frame, $h$ is  Planck's constant, $D=2.87$ GHz is the zero-field splitting in the NV, $\gamma_e=28$ GHz/T is the electron gyromagnetic ratio, and $\mathbf{B}(\an{x,y,}\:z) = \{B_x(\an{x,y,}\:z), B_y(\an{x,y,}\:z), B_z(\an{x,y,}\:z)\}$, denotes the inhomogeneous magnetic field generated by the external magnet. The force along $z$ due to a single spin is obtained as 
\dk{
\begin{equation} \label{eq:per_spin_force}
    f_z =\textbf{m}\cdot \partial_z {\bf B},
\end{equation}
where $\textbf{m}=-h\gamma_e\langle {\bf S}\rangle$ is the magnetic moment vector dependent on the NV spin state.}

This is analogous to the Stern-Gerlach experiment, where different spin states experience distinct forces in a magnetic gradient. We can calculate the total force along the $z$ direction on the oscillator due to the collective effect of the spins as  ${\bf{F}}=\int_V\,{\bf{f_z}}\,dV$, where the integration is over the volume of the diamond whose domain depends on the diamond's COM coordinates.

The \an{spin state of the} NV ensemble in a thermal state \an{is} described by $\rho_{\text{th}} = e^{-H_{\text{NV}}/k_\text{B} T}/Z$, $Z=\sum_{j=1}^3\, e^{-E_j/k_\text{B} T}
$, where $E_1, E_2, E_3$ are the eigenenergies of $H_{\text{NV}}$, $k_\text{B}$ is the Boltzmann constant, and $T=300\:\rm K$ at room temperature. 
The $z$-component of the effective magnetic moment per spin is averaged as $m_z =h\gamma_e\mathrm{Tr}(\rho_{\text{th}} S_z)$, generating a thermal force $f_{\text{th}}$, shown in Fig. \ref{fig:fig1} (C). We observe that $f_{\text{th}}$ depends on the relative angle $\theta$ between the NV axis and the local $\bf{B}$ field \an{at the spatial point $\{x,y,z\}$ of the NV center} and is directed towards higher $|\bf{B}|$.   

\an{To generate a mechanical drive due to the NV ensemble, we} periodically apply green laser pulses ($532 \:\rm nm$), with intensity $I$ and period the same as the mechanical period $T_\textrm{\dk{osc}}$.
The duty cycle of the laser is given by $D=t_{\text{GL}}/T_\textrm{\dk{osc}}$, where $t_{\text{GL}}$ is the on-time of the green light. 
This green light optically polarizes the NV spins. 
We assume that the time for the spins to polarize $t_{\text{pol}}<<T_\textrm{\dk{osc}}$. When the NV axis is aligned with the local $\bf B$ field ($\theta=0$), the optical spin polarization drives occupation mostly into the $m_s = 0$ spin state, reducing the overall spin force. However, when $\theta\ne 0$, the optical polarization can lead to a strong spin force due to spin-mixing.
This optical polarisation generates a net spin-force $F_{\text{GL}}$ over the entire diamond due to the inhomogeneous $\bf B$ field generated by the magnet interacting with the ensemble of NV centers in the diamond with four separate NV directions (see Supplementary Section 2). We illustrate the calculated per-spin optically generated spin-dependent force in Fig. \ref{fig:fig1} (C) as a function of $\theta$ and for various 532 nm laser illumination intensities (see the next section for more details on theoretical estimates of $F_{\text{GL}}$). We see in Fig. \ref{fig:fig1} (C) that substantial spin forces develop at certain angles $\theta$. After the laser is turned off, the spins relax back to $\rho_{\text{th}}$ within the longitudinal NV spin relaxation time $T_1$. Since both the polarization time $t_{\text{pol}}$ and $T_1$ are much shorter than $T_\textrm{\dk{osc}}$, the periodic laser pulses effectively switch the net spin force between $F_{\text{th}}$ and $F_{\text{GL}}$, thereby resonantly driving the oscillator.  
This driving scheme is illustrated in Fig. \ref{fig:fig1} (B), and, due to symmetry, \an{spin-dependent force}s along $x$- and $y$- directions cancel 
\dk{(see Supplementary Section 2)}. There is also a repulsive force between the resonator and the magnet due to the slight diamagnetism of the diamond material itself, but this is not spin-dependent, and we ignore this.
The periodic displacement of the resonator along $z$- is monitored to picoscale precision by an optical interferometer.

\dk{
\an{\subsection{Simulation of the spin force due to an NV ensemble}}
For simulating the per-spin force exerted on the NV center by the gradient magnetic field, expressed in Eq. \ref{eq:per_spin_force}, we need the externally applied magnetic field expression $\mathbf{B}(\mathbf{r})$ and the spin magnetic moment of the NV, $\mathbf{m}$.
The magnetic field generated by a cylindrical magnet was simulated using the exact analytical expression \cite{Caciagli2018ExactCylinder}.

The NV spin polarization depends on the intensity of the green light, the magnitude and orientation of the local magnetic field with respect to the NV quantization axis $\bf{B(\theta,\phi)}$, and the environmental temperature.
To predict the NV spin polarization due to pumping by green light, we use a seven-level rate equation model \cite{Tetienne2012SevenLevelNV, Patel2024SinglePhoto-excitation}. 
A seven level model calculates the spin population between the effective seven energy levels of NV centers, consisting of the NV's ground-state triplet, excited-state triplet, and a single intermediate state, using transition rates between them.
Using this model and solving the rate equation, we simulate the steady state population of the energy eigenstates, which we use to predict the resulting vector magnetic moment.
First, using the obtained steady state populations, we construct the density matrix in the energy eigenstate frame $\boldsymbol{\rho}_\textrm{e}$.
This density matrix representation is transformed to the NV spin projection basis through a basis transformation $\boldsymbol{\rho}_\textrm{NV} = \boldsymbol{U}^T \boldsymbol{\rho}_\textrm{e} \boldsymbol{U}$, where $\boldsymbol{U}$ is the unitary matrix constructed by the energy eigenvectors.
Then, we calculate the spin polarization expectation value using $\left<S_i \right>_\textrm{NV} = \textrm{Tr}[\rho S^\textrm{NV}_i]$ to obtain $\left< \mathbf{S}\right>_\textrm{NV}=\{\left< S_x\right>_\textrm{NV},\left<S_y\right>_\textrm{NV}, \left< S_z\right>_\textrm{NV}\}$.
The spin expectation value vector is then rotated to the laboratory frame using a rotation matrix $R_i$, which is unique to each NV orientation,
$
\left<\mathbf{S} \right>_\textrm{lab} = 
\boldsymbol{R}\left<\mathbf{S} \right>_\textrm{NV}.
$
The rotation matrix $\boldsymbol{R}_i=\boldsymbol{R}_z(\phi_i)\boldsymbol{R}_y(\theta_i)$ maps vector components from the NV frame to the laboratory frame for the $i$th NV orientation, where the angles $\theta_i$ and $\phi_i$ specify the polar and azimuthal angles, respectively, of the NV symmetry axis expressed in the laboratory frame.
To account for the four orientations of NV centers with equal probability distribution, we calculate all four directions at the position of interest and average the force by a factor of 4.

To simulate the spin-dependent force on the hybrid levitated oscillator, we integrate the force densities within the illuminated volume of the diamond to obtain the net force, assuming the induced magnetic moment is uniform over the laser spot size of $\sim 1$ mm, with the magnetic moment chosen to be at the center of the laser irradiation spot where the laser intensity is maximum.
Since we are alternating the population between thermal and optically polarized states, we calculate the difference in force.
\an{We scale this force down by a scaling factor of \annew{$1.2$}, consistent for all our measurements.}
Exact details of the simulations for the optical spin polarisation and the force can be found in Supplementary Section 2.
}

\subsection{Observation of spin force}

\begin{figure} 
	\centering
	
    \includegraphics[width=0.9\textwidth]{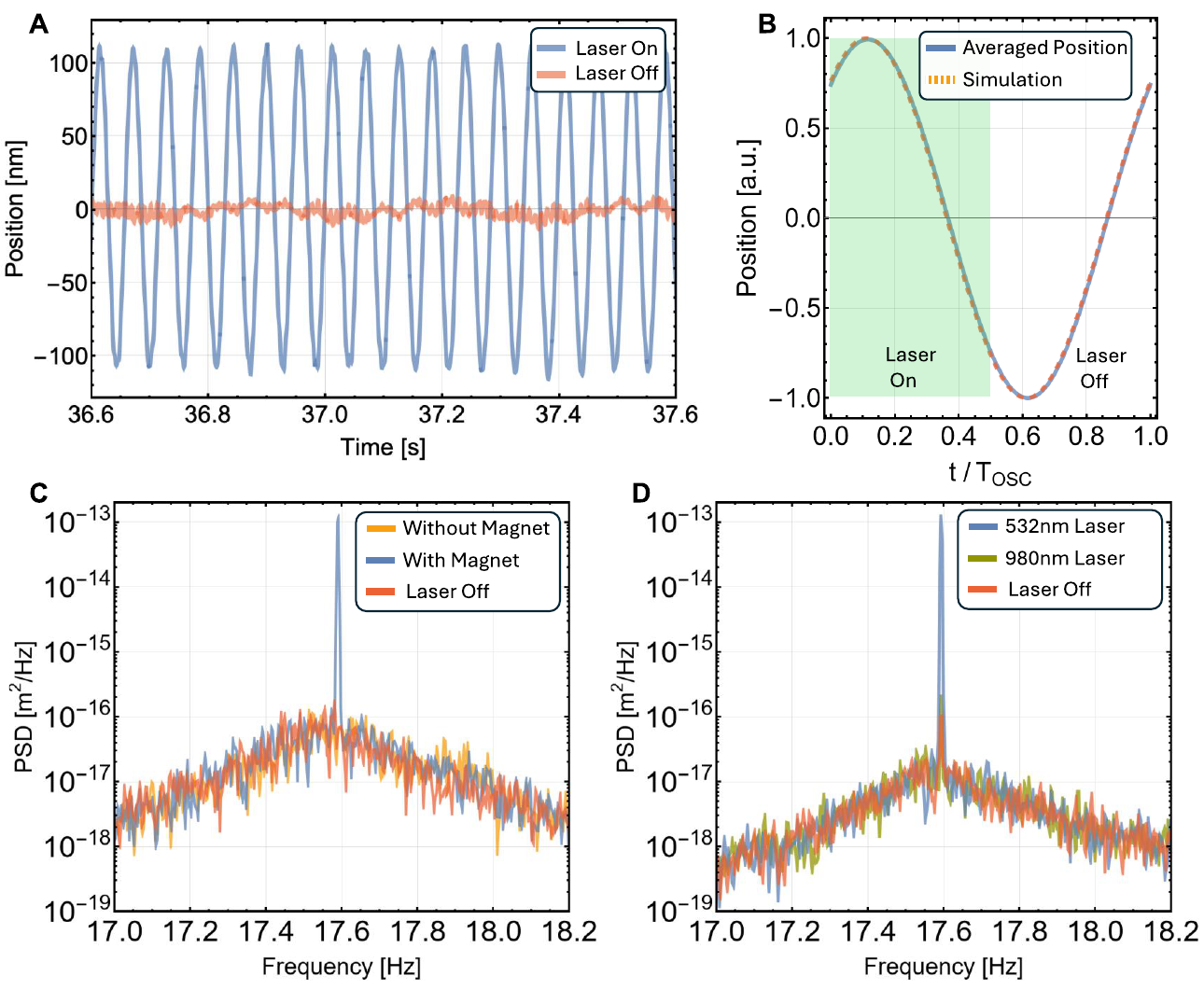}
    
	\caption{ \textbf{Driving the center of mass of \dk{a levitated} oscillator \dk{via optical} spin polarisation:} 
\textbf{(A)} Orange line shows the levitated oscillator's Brownian motion at atmospheric pressure. Periodic 532 nm illumination of NV centers in the diamond, near the oscillator resonance (17.6 Hz) with 48.73 duty cycle and 50 mW power, modulates the NV spin population (and thus its magnetic dipole moment), driving the center-of-mass motion above Brownian noise. 
\textbf{(B)} Oscillator motion for one mechanical period, obtained by averaging over 1200 s with a moving window of $56.8\:\rm ms$ (oscillation period), reveals persistent spin-driven oscillations consistent with simulations of a Brownian harmonic oscillator driven by a square-wave force.
\textbf{(C)} Power spectral density under 10 mW periodic illumination with and without a bias magnet. The magnet provides both the magnetic gradient for \an{spin-dependent force} and Zeeman splitting for differential spin population. A sharp resonance peak appears with the magnet and disappears without it. 
\textbf{(D)} As control, a 980 nm IR laser (50 mW), which does not polarise NV spins, is pulsed with a 50\% duty cycle. The same pulsing applied to the 532 nm laser (10 mW) produces a clear resonance peak, confirming NV-mediated driving. The small residual peak without green light originates from chopper noise \an{which has been kept switched on for all the measurements}. }
	\label{fig:fig2} 
\end{figure}

We first test our protocol by resonantly pulsing the 532 nm laser set at 50 mW power and 48\% duty cycle, to observe the oscillator's motion due to the spin-mechanical coupling. The periodic \an{spin-dependent force} on the levitated oscillator should shift its equilibrium position, and the motion should subsequently be amplified by the oscillator's moderately high quality factor $Q_\text{m}$. When the protocol is experimentally performed, we observe in real-time a strong COM oscillation with a steady-state amplitude of $\sim 100$~nm, which is \an{substantially} larger than the nanometer-scale Brownian motion, as shown in Fig. \ref{fig:fig2} (A). 
\dk{
We divide the time trace into consecutive segments, each equal to one mechanical oscillation period, and average them (Fig. \ref{fig:fig2} (B)). The resulting waveform is then normalized in amplitude. If the motion lacked phase coherence, for example, because of Brownian fluctuations, phase drift, or frequency variations, the oscillation would be reduced by this averaging. Instead, a clear sinusoidal signal remains and agrees with the expected response to the periodic rectangular drive. This shows that the motion is phase-coherent.}
\an{The sinusoidal nature of the averaged motion in Fig. \ref{fig:fig2} (B) indicates that the dynamics of the hybrid oscillator under this large amplitude drive can still be well approximated by that of a harmonic oscillator with negligible nonlinearities.} We observe that when the laser is on, the motion of the oscillator moves towards the magnet, i.e., towards $-z$, indicative that the force is attractive, which is predicted by the theoretical model. 

Our design aims to keep the diamond slab away from any nearby surfaces and surround the diamond with unobstructed access to the air in order to reduce any heat-induced gas forces, known as photophoretic forces, as the laser irradiation can elevate the diamond's bulk temperature by $\sim 3$K \an{(See Supplementary Section 8)}. The inclusion of the iron plate also shields the laser interferometer displacement sensor from the 532 nm laser, eliminating the potential for any cross-talk. However, for the modulation of the 980 nm laser, we use a mechanical chopper for the pulse modulation. We notice a small \an{mechanical} excitation of the resonator's motion when the laser is off and the chopper is on, even though the chopper is positioned \an{ outside the vibration isolation table}. The modulation of the 532 nm laser is electronic, and no sympathetic position modulation is observed when the laser is off.

To verify that these oscillations arise from spin-mechanical interactions, we perform additional experiments: firstly, we pulse a 10 mW green laser in identical setups except that the permanent magnet supplying the magnetic field gradient to the diamond slab is either present or not.   Secondly, we use a setup where the permanent magnet is present, but replace the 532 nm laser with a  50 mW 980 nm infrared (IR) laser. This long wavelength is incapable of optical polarization of the NV spins and thus, no modulation of the NV spin states or diamond slab position should occur when the IR laser radiation is modulated in a rectangular pulse. 
A higher IR laser power is chosen to account for the fact that the diamond is mostly transparent to the IR laser. Fig. \ref{fig:fig2} (C) and (D) show the motional power spectral density (PSD) under these separate setups. In the presence of both the magnet and 532 nm polarizing laser, a pronounced, sharp peak with roughly 4 orders of magnitude over the Brownian noise appears in the PSD at the COM resonance, corresponding to strong coherent motion. In contrast, when the magnet is absent (Fig. \ref{fig:fig2} (C)), or when a non-polarizing laser is used (Fig. \ref{fig:fig2} (D)), this strong motional resonance peak disappears. These results demonstrate that the driving force is magnetic and depends on the spin polarization of the NV centers, confirming that the mechanical motion is mediated by the spin, rather than by thermal effects or optical artefacts. \an{The amplitude of the oscillator is limited by the damping rate due to air resistance. On moving to high vacuum, we can reduce the damping rate due to air and get motional amplitude in the order of micrometers (see Supplementary Section 6).}

\subsection{Quantifying the \an{spin-dependent force} from the motional Power Spectral Density}

\begin{figure} 
	\centering
	\includegraphics[width=1\textwidth]{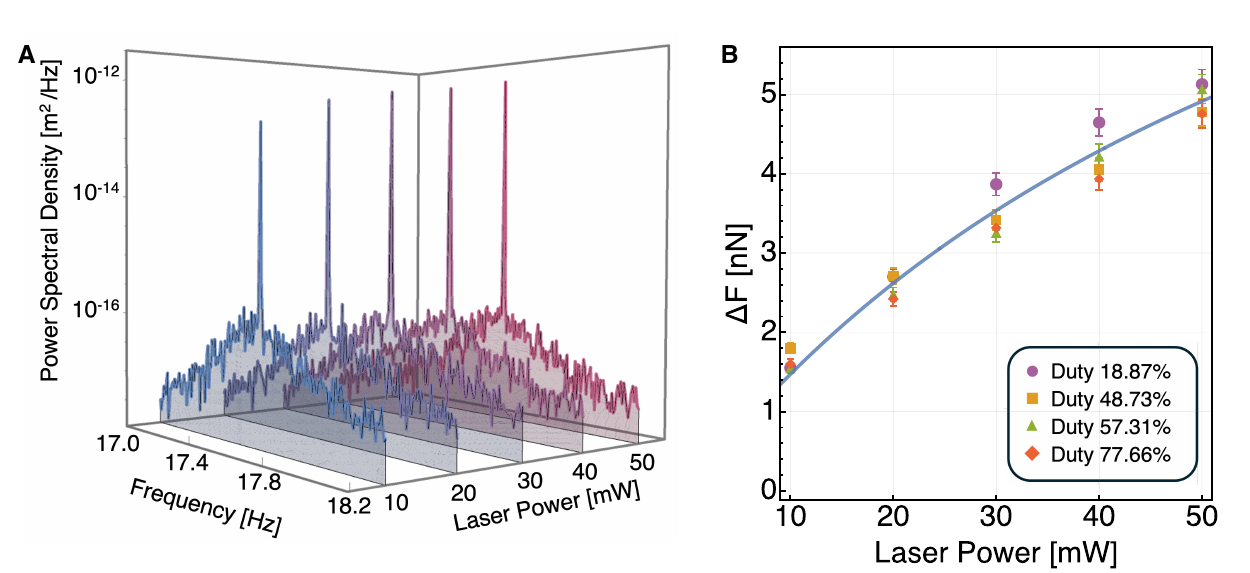} 
	\caption{\textbf{Measuring and controlling the spin force \dk{ via laser power and duty cycle}.}
		\textbf{(A)} \an{Waterfall plot of PSD around resonance frequency due to laser drive at 48$\%$ duty cycle for various laser powers. As we increase the laser power, we can observe the peak due to the force in the PSD is increasing.} \textbf{(B)} Estimated spin force {\color{black}amplitude} from the displacement amplitude data as a function of laser power. The experimental data (points) show a good fit with our prediction obtained from the seven-level spin model (solid line), and predict that the spin force increases with an increase in laser power. The motion of the oscillator can be controlled arbitrarily by tuning the laser power and the duty cycle of the pulse. }
	\label{fig:fig3} 
\end{figure}

We now estimate the magnitude of the \an{spin-dependent force} that gives rise to the experimentally observed motional PSDs. This estimate will be compared later to numerical simulations of the net spin-force. As the 532 nm laser drive is periodic rectangular with frequency $f_0$ (matching the mechanical resonance), the resultant force profile is also rectangular. We can vary the duty cycle $D$ of the laser pulses, which will also alter the overall motional oscillation amplitude.  
We position the hanging diamond slab to have a nominal gap separation of $d=0.5$ mm to the permanent magnet below and record the resonator's motion when the laser drive is set to have various duty cycles, keeping the peak power the same. After some initial transient time, the resonator will exhibit periodic sinusoidal motion with amplitude $A(f_0,D)$. To extract amplitude information $A(f_0,D)$, we first note that since the drive is periodic, the resulting motional PSD $S_{xx}$ consists of delta functions in $\omega$ at the harmonics of the driving frequency (see Supplementary Material). With finite measurement time and resolution, these delta functions manifest as sharp peaks in the numerically estimated PSD from the measured displacement signal, whose peak heights increase and widths decrease, with increasing measurement time - approximate delta functions. Since $\int_{\omega_0-\epsilon}^{\omega_0+\epsilon}\,d\omega\,\delta(\omega-\omega_0)=1$, we can integrate the area under the primary resonance peak in the PSD, $S_{xx}$, to obtain the estimate of the oscillation amplitude via
    $A(f_0,D)^2=\int_{f_0-\delta f}^{f_0+\delta f} S_{xx}(f) df$\;\;.
We note that the amplitude $A$ also depends on the intensity of the laser pulse, as the induced spin polarization will vary with intensity. As we approach 50\% duty cycle, we get stronger oscillations of the center of mass, for all laser powers, which then subsequently decreases as we further increase the duty cycle, as seen in the density plot Fig. \ref{fig:fig3} (A). We obtain the force amplitude $\Delta F$ from the displacement amplitude by 
\begin{equation}
\Delta F= \frac{\pi}{2}\frac{m(2\pi f_0)^2}{Q_\text{m} \sin(\pi D)}A(f_0,D)\;\;,\label{eq:force}
\end{equation}
where $m$ is the mass of the resonator (see Supplementary Section 3).

We plot the magnitude of the force derived by Eq. \ref{eq:force}, from the measurements as a function of power for various duty cycles in Fig. \ref{fig:fig3} (B) [points with error bars], and the simulated force estimate [solid line]. We observe that our numerically simulated predicted force agrees well for all the duty cycles and increases with increasing laser power, approaching a saturation value. This trend is captured by our simulations based on the seven-level model. 

\subsection{\dk{Magnetic field gradient} dependence of spin force}
\begin{figure} 
	\centering
	\includegraphics[width=1\textwidth]{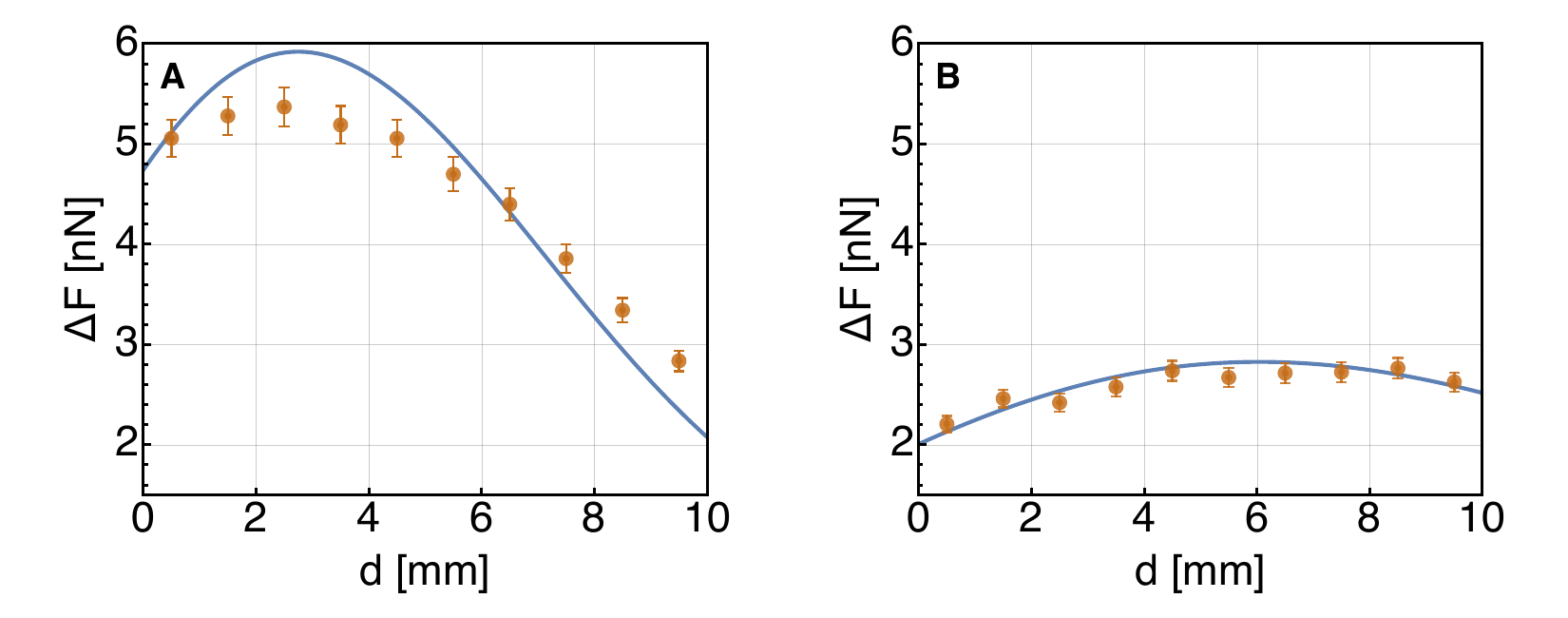} 
	\caption{\textbf{\dk{Magnetic field gradient} dependent force:} 
    We vary the gap separation $d$ between the diamond and the cylinder magnet by lowering the magnet relative to the diamond. This keeps the diamond and laser irradiation spot on the diamond unchanged. For each value of $d$, we drive the spin-mechanical motion using a rectangular laser pulse, and estimate the amplitude of the \an{spin-dependent force} from the motional PSD. The smooth curve is a numerical simulation using the seven-level NV polarization model, where we assume the induced spin polarization is uniform over the laser irradiation volume in the diamond. \textbf{(A)} The results shown using a small permanent magnet and 
    \textbf{(B)} the spin force using a larger magnet that produces a stronger field but a lower gradient, indicating that this force is sensitive to the gradient and not to the magnitude of $\bf B$.}
	\label{fig:fig4} 
\end{figure}

Finally,  we investigate how the amplitude of the \an{spin-dependent force} varies as we vary the gap distance $d$ between the bias magnet and the diamond. The distance is controlled using a translation stage, with 
$d$ increased in steps of 1 mm, which reduces the magnitude of both the magnetic field $\bf{B}$ and the magnetic gradient $G_z$. At each height, we perform the resonant laser driving scheme with 50 mW laser power and 48\% duty cycle. We repeated these measurements with two different cylindrical magnets, the second one twice the diameter of the first and each having similar magnetizations. The second magnet produces a lower magnetic gradient in the location of the diamond \an{(See Supplementary Section 1)}. The experimental results of $\Delta F(d)$ are shown in Fig. \ref{fig:fig4} (A) and (B) for these two magnets.

We note that the force measured with the first magnet is stronger than with the second, 
\dk{even in the region where the latter produces a larger magnetic field.}
This supports that the motion arises from the spins coupling to the magnetic gradient, and the motion does not arise from a spin-torque where the spins couple to the B-field itself.

\section{Discussion}
To summarize, we have demonstrated the motional driving of a massive diamagnetically levitated mechanical oscillator by the force of NV spins in a gradient magnetic field. \an{This result demonstrates spin-mechanical coupling in a massive, sub-gram, centimeter-scale regime and constitutes a direct measurement of the magnetic-gradient-induced force on an oscillator generated by the NV centers.} Our experiments operate with a modest magnetic gradient in the range of 10 T/m to 100 T/m generated by a single permanent magnet, and through that we measure forces \an{reaching over 5 nN} due to a change in spin populations much smaller than a complete state transfer. These results show that even a small change in the NV spin state can give rise to a dramatic mechanical response, making the system quite attractive for sensing weak signals. \an{A key point to note is that the sensitivity to measure the position in our measurements is not limited by the interferometric setup, but by the ambient noise in the lab, which sets the noise in our measurements to about 1 nm. Improving the vibrational isolation around our setup can enable us to improve the measurement sensitivity by 3 orders of magnitude to the sensor's noise floor (in picometer range).}

Our resonator design spatially separates the diamond from the levitating element, with which we can avoid detrimental effects sometimes associated with heating forces (photophoretic), even when in high vacuum.  Assuming that the optically induced moment is equal to the polarization at the laser center, our simulations show excellent agreement with the experimental results. \an{Moving the setup to high vacuum would enhance the amplitude of the motion by at least an order of magnitude, still limited by eddy damping in the graphite oscillator.} Combined with high vacuum, replacing the slotted graphite resonator with a composite diamagnetic resonator would enable a motional $Q$-factors $\sim 10^5$ \cite{Chen2022DiamagneticResonators, Tian2024FeedbackPlate} corresponding to a large motional amplitude in the order of 100 $\mu \rm m$, with just the weak thermal spin polarisation. Further, advances in microwave control of large ensembles of spins will enable us to gain coherent control over the spin states of NV and full state transfer of the spins, potentially giving rise to a stronger force. \an{By employing coherent microwave control techniques and shorter laser pulses, which would cause lower heating of the diamond, the spurious effects on the mechanical motion arising from laser-induced heating of the oscillator can be substantially mitigated.}
One could then envision engineering coherent energy exchanges between spin and mechanics, with spin-cooling of the COM in the near term and, eventually, tests of macroscopic non-classical states of motion, novel spin-mechanical interactions, and quantum thermodynamics.

\an{\section{Materials and Methods}}
\an{
The levitated system to demonstrate spin-mechanical coupling comprises a bulk diamond slab, which is oriented so that its face normals are horizontal, and hangs from a diamagnetically levitated slotted graphite slab via a thin rigid carbon fibre rod.

A commercial single-crystal diamond slab with uniformly distributed NV centers  (THORLABS DNVB14) is used in the experiment. 
The diamond slab has dimensions $3\rm\:mm \times 3\:mm\times0.5\:mm$ with an NV center concentration of 4.5 ppm. All the faces of the diamonds are cut along [1,0,0] direction, with the four groups of NV making $\sim54.7^{\circ}$ with the vertical \cite{Yang2022NVOrientation}.

The diamond is attached to the bottom of a thin, rigid carbon fiber rod of length 5 cm and diameter 0.5 mm. The top of the rod is attached to the center bottom of a pyrolytic graphite plate of dimensions $10\rm\:mm \times 10\:mm\times1\:mm$.   These attachments are made using UV glue.
The graphite plate is levitated above four N52 NdFeB magnets with magnetization in a checkerboard array, which forms the diamagnetic trap. The cubic magnet has a side length of $12.7\rm\:mm$. Circular-patterned slots are cut through the graphite plate using a femtosecond laser to suppress the eddy currents and mitigate the mechanical damping. 
Details of the magnetic levitation setup and eddy damping are presented in \cite{Romagnoli2023ControllingPlate}. 
A small hole is drilled at the center of the magnet array to accommodate the passage of the carbon fiber rod. 
A small mirror is glued onto the center of the top surface of the graphite plate for interferometric measurement.
A fiber-optic head of a commercial interferometer (SmarAct Picoscale) is mounted above the mirror to bounce light off the mirror and measure position in time via interference to picometer-level precision. The setup is placed on an active vibrational isolation table (Table Stable TS-300) and shielded from air currents by a clear acrylic box.

We place an iron plate (4 mm thick, with high magnetic permeability) under the checkerboard magnet array to shield the diamond slab from the magnetic field of the checkerboard array. A small hole is drilled in this plate to admit the thin carbon-fiber rod. 

The diamond is illuminated by a 532 nm  green laser (Edmund Optics MGL-III-532-50mW-M), \annew{whose polarisation fluctates $\pm5$ degrees over time}, which is pulsed by an external trigger generated by a digital oscilloscope (Cleverscope CS328A). For the controlled experiment in which we illuminate the diamond with light whose wavelength is incapable of spin polarization, we use a 980 nm IR laser (50 mW, Amazon 980MD-200-12V). This wavelength is far beyond the zero phonon line $637$ nm of NV${}^-$ in diamond. This IR laser does not have external modulation capabilities, and hence, we modulate the diamond illuminated using a precision mechanical chopper (Newport 3502). The measurements are done at room temperature, 300 K, and atmospheric pressure, with position data $z(t)$ being collected at a sampling rate of 2440 Hz.

Under the diamond plate, we position a cylindrical permanent N52 magnet on an XYZ-stage. We use two sizes of magnets whose dimensions are given in Table S1. \an{The simulations of the magnetic fields due to these magnets were obtained through simulations of the exact analytical expression following the reference \cite{Caciagli2018ExactCylinder} and are provided in Supplementary Section 1.
}
}

\clearpage 

\bibliographystyle{ScienceAdvances}

\begin{thebibliography}{10}

\bibitem{Gerlach1922DerMagnetfeld}
W.~Gerlach, O.~Stern, {Der experimentelle Nachweis der Richtungsquantelung im Magnetfeld}.
\newblock {\it Zeitschrift fur Physik\/} {\bf 9}, 349--352 (1922).

\bibitem{Klabunde1934TheIron}
W.~Klabunde, T.~E. Phipps, {The Stern-Gerlach Experiment with Iron}.
\newblock {\it Physical Review\/} {\bf 45}, 59--61 (1934).

\bibitem{Estermann1937TheProton}
I.~Estermann, O.~C. Simpson, O.~Stern, {The Magnetic Moment of the Proton}.
\newblock {\it Physical Review\/} {\bf 52}, 535--545 (1937).

\bibitem{Koppe1948DasElektrons}
H.~Koppe, {Das magnetische Moment des Elektrons}.
\newblock {\it Zeitschrift fur Naturforschung - Section A Journal of Physical Sciences\/} {\bf 3}, 124--125 (1948).

\bibitem{Sherwood1954Stern-GerlachNeutrons}
J.~E. Sherwood, T.~E. Stephenson, S.~Bernstein, {Stern-Gerlach Experiment on Polarized Neutrons}.
\newblock {\it Physical Review\/} {\bf 96}, 1546--1548 (1954).

\bibitem{Bucher1991MagneticBeam}
J.~P. Bucher, D.~C. Douglass, P.~Xia, B.~Haynes, L.~A. Bloomfield, {Magnetic deflection of neutral metal clusters in a beam}.
\newblock {\it Zeitschrift f{\"{u}}r Physik D Atoms, Molecules and Clusters\/} {\bf 19}, 251--254 (1991).

\bibitem{Douglass1993MagneticClusters}
D.~C. Douglass, A.~J. Cox, J.~P. Bucher, L.~A. Bloomfield, {Magnetic properties of free cobalt and gadolinium clusters}.
\newblock {\it Physical Review B\/} {\bf 47}, 12874--12889 (1993).

\bibitem{Robert1991AtomicAtoms}
J.~Robert, C.~Miniatura, S.~L. Boiteux, J.~Reinhardt, V.~Bocvarski, J.~Baudon, {Atomic Interferometry with Metastable Hydrogen Atoms}.
\newblock {\it Europhysics Letters (EPL)\/} {\bf 16}, 29--34 (1991).

\bibitem{Rabi1988OttoQuantization}
I.~I. Rabi, J.~S. Rigden, {Otto Stern and the discovery of space quantization}.
\newblock {\it Zeitschrift f{\"{u}}r Physik D Atoms, Molecules and Clusters\/} {\bf 10}, 119--120 (1988).

\bibitem{MacHluf2013CoherentChip}
S.~Machluf, Y.~Japha, R.~Folman, {Coherent Stern–Gerlach momentum splitting on an atom chip}.
\newblock {\it Nature Communications\/} {\bf 4}, 2424 (2013).

\bibitem{Wu2019SternGerlachLattice}
T.~Y. Wu, A.~Kumar, F.~Giraldo, D.~S. Weiss, {Stern–Gerlach detection of neutral-atom qubits in a state-dependent optical lattice}.
\newblock {\it Nature Physics\/} {\bf 15}, 538--542 (2019).

\bibitem{Amit2019T3Interferometer}
O.~Amit, Y.~Margalit, O.~Dobkowski, Z.~Zhou, Y.~Japha, M.~Zimmermann, M.~A. Efremov, F.~A. Narducci, E.~M. Rasel, W.~P. Schleich, R.~Folman, {T3 Stern-Gerlach Matter-Wave Interferometer}.
\newblock {\it Physical Review Letters\/} {\bf 123}, 83601 (2019).

\bibitem{Rabl2009StrongResonator}
P.~Rabl, P.~Cappellaro, M.~V.~G. Dutt, L.~Jiang, J.~R. Maze, M.~D. Lukin, {Strong magnetic coupling between an electronic spin qubit and a mechanical resonator}.
\newblock {\it Physical Review B\/} {\bf 79}, 041302 (2009).

\bibitem{Abdi2017hBN}
M.~Abdi, M.-J. Hwang, M.~Aghtar, M.~B. Plenio, Spin-mechanical scheme with color centers in hexagonal boron nitride membranes.
\newblock {\it Phys. Rev. Lett.\/} {\bf 119}, 233602 (2017).

\bibitem{Teissier2014StrainOscillator}
J.~Teissier, A.~Barfuss, P.~Appel, E.~Neu, P.~Maletinsky, {Strain Coupling of a Nitrogen-Vacancy Center Spin to a Diamond Mechanical Oscillator}.
\newblock {\it Physical Review Letters\/} {\bf 113}, 020503 (2014).

\bibitem{Ovartchaiyapong2014DynamicResonator}
P.~Ovartchaiyapong, K.~W. Lee, B.~A. Myers, A.~C. Jayich, {Dynamic strain-mediated coupling of a single diamond spin to a mechanical resonator}.
\newblock {\it Nature Communications\/} {\bf 5}, 6--11 (2014).

\bibitem{Fedele2024CouplingMotion}
F.~Fedele, F.~Cerisola, L.~Bresque, F.~Vigneau, J.~Monsel, J.~Tabanera, K.~Aggarwal, J.~Dexter, S.~Sevitz, J.~Dunlop, A.~Auff{\`{e}}ves, J.~Parrondo, A.~P{\'{a}}lyi, J.~Anders, N.~Ares, {Coupling a single spin to high-frequency motion}.
\newblock {\it arxiv:2402.19288\/} pp. 1--13 (2024).

\bibitem{Wallquist2009HybridEngineering}
M.~Wallquist, K.~Hammerer, P.~Rabl, M.~Lukin, P.~Zoller, {Hybrid quantum devices and quantum engineering}.
\newblock {\it Physica Scripta\/} {\bf T137}, 014001 (2009).

\bibitem{Mamin2004SingleMicroscopy}
H.~J. Mamin, D.~Rugar, B.~W. Chui, R.~Budakian, {Single spin detection by magnetic resonance force microscopy}.
\newblock {\it Nature\/} {\bf 430}, 329--332 (2004).

\bibitem{ShimonKolkowitz2012CoherentQubit}
{Shimon Kolkowitz}, {Ania C. Bleszynski Jayich}, {Quirin P. Unterreithmeier}, {Steven D. Bennett}, {Peter Rabl}, {J. G. E. Harris}, {Mikhail D. Lukin}, {Coherent Sensing of a MechanicalResonator with a Single-Spin Qubit}.
\newblock {\it Science\/} {\bf 335}, 1600--1603 (2012).

\bibitem{Hong2012CoherentSpin}
S.~Hong, M.~S. Grinolds, P.~Maletinsky, R.~L. Walsworth, M.~D. Lukin, A.~Yacoby, {Coherent, mechanical control of a single electronic spin}.
\newblock {\it Nano Letters\/} {\bf 12}, 3920--3924 (2012).

\bibitem{Chotorlishvili2013EntanglementResonator}
L.~Chotorlishvili, D.~Sander, A.~Sukhov, V.~Dugaev, V.~R. Vieira, A.~Komnik, J.~Berakdar, {Entanglement between nitrogen vacancy spins in diamond controlled by a nanomechanical resonator}.
\newblock {\it Physical Review B\/} {\bf 88}, 085201 (2013).

\bibitem{Li2015HybridCavities}
P.-B. Li, Y.-C. Liu, S.-Y. Gao, Z.-L. Xiang, P.~Rabl, Y.-F. Xiao, F.-L. Li, {Hybrid Quantum Device Based on NV Centers in Diamond Nanomechanical Resonators Plus Superconducting Waveguide Cavities}.
\newblock {\it Physical Review Applied\/} {\bf 4}, 044003 (2015).

\bibitem{Xia2016}
K.~Xia, J.~Twamley, {Generating spin squeezing states and Greenberger-Horne-Zeilinger entanglement using a hybrid phonon-spin ensemble in diamond}.
\newblock {\it Physical Review B\/} {\bf 94}, 205118 (2016).

\bibitem{Zhang2021QuantumNanodiamond}
H.~Zhang, X.~Chen, Z.-q. Yin, {Quantum Information Processing and Precision Measurement Using a Levitated Nanodiamond}.
\newblock {\it Advanced Quantum Technologies\/} {\bf 4}, 1--12 (2021).

\bibitem{Zhao2022InertialDiamond}
L.~Zhao, X.~Shen, L.~Ji, P.~Huang, {Inertial measurement with solid-state spins of nitrogen-vacancy center in diamond}.
\newblock {\it Advances in Physics: X\/} {\bf 7}, 2004921 (2022).

\bibitem{Fung2024TowardTransport}
F.~Fung, E.~Rosenfeld, J.~D. Schaefer, A.~Kabcenell, J.~Gieseler, T.~X. Zhou, T.~Madhavan, N.~Aslam, A.~Yacoby, M.~D. Lukin, {Toward Programmable Quantum Processors Based on Spin Qubits with Mechanically Mediated Interactions and Transport}.
\newblock {\it Physical Review Letters\/} {\bf 132}, 263602 (2024).

\bibitem{Arrazola2024TowardPhonons}
I.~Arrazola, Y.~Minoguchi, M.-A. Lemonde, A.~Sipahigil, P.~Rabl, {Toward high-fidelity quantum information processing and quantum simulation with spin qubits and phonons}.
\newblock {\it Physical Review B\/} {\bf 110}, 045419 (2024).

\bibitem{Chia2024HybridState}
C.~Chia, D.~Huang, V.~Leong, J.~F. Kong, K.~E.~J. Goh, {Hybrid Quantum Systems with Artificial Atoms in Solid State}.
\newblock {\it Advanced Quantum Technologies\/} {\bf 7}, 1--21 (2024).

\bibitem{Yin2013}
Z.-Q. Yin, T.~Li, X.~Zhang, L.~M. Duan, {Large quantum superpositions of a levitated nanodiamond through spin-optomechanical coupling}.
\newblock {\it Physical Review A\/} {\bf 88}, 033614 (2013).

\bibitem{Scala2013}
M.~Scala, M.~S. Kim, G.~W. Morley, P.~F. Barker, S.~Bose, {Matter-Wave Interferometry of a Levitated Thermal Nano-Oscillator Induced and Probed by a Spin}.
\newblock {\it Physical Review Letters\/} {\bf 111}, 180403 (2013).

\bibitem{Bose2017}
S.~Bose, A.~Mazumdar, G.~W. Morley, H.~Ulbricht, M.~Toro{\v{s}}, M.~Paternostro, A.~A. Geraci, P.~F. Barker, M.~S. Kim, G.~Milburn, {Spin Entanglement Witness for Quantum Gravity}.
\newblock {\it Physical Review Letters\/} {\bf 119}, 240401 (2017).

\bibitem{Zhou2025Spin-dependentSuperpositions}
R.~Zhou, Q.~Xiang, A.~Mazumdar, {Spin-dependent force and inverted harmonic potential for rapid creation of macroscopic quantum superpositions}.
\newblock {\it Physical Review A\/} {\bf 111}, 52207 (2025).

\bibitem{RamanNair2025MassiveMagnetomechanics}
S.~Raman~Nair, S.~Tian, G.~K. Brennen, S.~Bose, J.~Twamley, {Massive quantum superpositions using magnetomechanics}.
\newblock {\it Physical Review Applied\/} {\bf 24}, 024061 (2025).

\bibitem{Gonzalez-Ballestero2021Levitodynamics:Vacuum}
C.~Gonzalez-Ballestero, M.~Aspelmeyer, L.~Novotny, R.~Quidant, O.~Romero-Isart, {Levitodynamics: Levitation and control of microscopic objects in vacuum}.
\newblock {\it Science\/} {\bf 374}, 168 (2021).

\bibitem{Marletto2017}
C.~Marletto, V.~Vedral, {Gravitationally Induced Entanglement between Two Massive Particles is Sufficient Evidence of Quantum Effects in Gravity}.
\newblock {\it Physical Review Letters\/} {\bf 119}, 240042 (2017).

\bibitem{Bose2025MassiveGravity}
S.~Bose, I.~Fuentes, A.~A. Geraci, S.~M. Khan, S.~Qvarfort, M.~Rademacher, M.~Rashid, M.~Toro{\v{s}}, H.~Ulbricht, C.~C. Wanjura, {Massive quantum systems as interfaces of quantum mechanics and gravity}.
\newblock {\it Reviews of Modern Physics\/} {\bf 97}, 015003 (2025).

\bibitem{Qin2019ProposalResonators}
W.~Qin, A.~Miranowicz, G.~Long, J.~Q. You, F.~Nori, {Proposal to test quantum wave-particle superposition on massive mechanical resonators}.
\newblock {\it npj Quantum Information\/} {\bf 5}, 58 (2019).

\bibitem{Sahoo2023EmergenceSelfGravity}
S.~K. Sahoo, R.~Vathsan, T.~Qureshi, {Emergence of Classicality in Stern–Gerlach Experiment via Self‐Gravity}.
\newblock {\it Annalen der Physik\/} {\bf 535}, 1--6 (2023).

\bibitem{Grossardt2024Self-gravitationalTrajectories}
A.~Grossardt, {Self-gravitational dephasing of quasiclassical Stern-Gerlach trajectories}.
\newblock {\it Physical Review D\/} {\bf 109}, 1--5 (2024).

\bibitem{Carney2020ProposalMatter}
D.~Carney, S.~Ghosh, G.~Krnjaic, J.~M. Taylor, {Proposal for gravitational direct detection of dark matter}.
\newblock {\it Physical Review D\/} {\bf 102}, 72003 (2020).

\bibitem{Monteiro2020a}
F.~Monteiro, G.~Afek, D.~Carney, G.~Krnjaic, J.~Wang, D.~C. Moore, {Search for Composite Dark Matter with Optically Levitated Sensors}.
\newblock {\it Physical Review Letters\/} {\bf 125}, 181102 (2020).

\bibitem{Figueroa2021DarkNetworks}
N.~L. Figueroa, D.~Budker, E.~M. Rasel, {Dark matter searches using accelerometer-based networks}.
\newblock {\it Quantum Science and Technology\/} {\bf 6}, 034004 (2021).

\bibitem{Zhang2014}
X.~Zhang, C.-L. Zou, L.~Jiang, H.~X. Tang, Strongly coupled magnons and cavity microwave photons.
\newblock {\it Physical Review Letters\/} {\bf 113}, 156401 (2014).

\bibitem{Neukirch2015Multi-dimensionalNanodiamond}
L.~P. Neukirch, E.~von Haartman, J.~M. Rosenholm, A.~Nick~Vamivakas, {Multi-dimensional single-spin nano-optomechanics with a levitated nanodiamond}.
\newblock {\it Nature Photonics\/} {\bf 9}, 653--657 (2015).

\bibitem{Hoang2016ElectronVacuum}
T.~M. Hoang, J.~Ahn, J.~Bang, T.~Li, {Electron spin control of optically levitated nanodiamonds in vacuum}.
\newblock {\it Nature Communications\/} {\bf 7}, 12250 (2016).

\bibitem{Delord2020}
T.~Delord, P.~Huillery, L.~Nicolas, G.~H{\'{e}}tet, {Spin-cooling of the motion of a trapped diamond}.
\newblock {\it Nature\/} {\bf 580}, 56--59 (2020).

\bibitem{Pellet-Mary2021MagneticInteractions}
C.~Pellet-Mary, P.~Huillery, M.~Perdriat, G.~H{\'{e}}tet, {Magnetic torque enhanced by tunable dipolar interactions}.
\newblock {\it Physical Review B\/} {\bf 104}, L100411 (2021).

\bibitem{Perdriat2021Spin-mechanicsParticles}
M.~Perdriat, C.~Pellet-Mary, P.~Huillery, L.~Rondin, G.~H{\'{e}}tet, {Spin-Mechanics with Nitrogen-Vacancy Centers and Trapped Particles}.
\newblock {\it Micromachines\/} {\bf 12}, 651 (2021).

\bibitem{Perdriat2025Spin-dependentMicrolever}
M.~Perdriat, A.~Durand, L.~Chambard, J.~Voisin, G.~H{\'{e}}tet, {Spin-dependent force from an NV center ensemble on a microlever}.
\newblock {\it Physical Review B\/} {\bf 112}, 094419 (2025).

\bibitem{Caciagli2018ExactCylinder}
A.~Caciagli, R.~J. Baars, A.~P. Philipse, B.~W.~M. Kuipers, Exact expression for the magnetic field of a finite cylinder with arbitrary uniform magnetization.
\newblock {\it Journal of Magnetism and Magnetic Materials\/} {\bf 456}, 423--432 (2018).

\bibitem{Tetienne2012SevenLevelNV}
J.-P. Tetienne, L.~Rondin, P.~Spinicelli, M.~Chipaux, T.~Debuisschert, J.-F. Roch, V.~Jacques, Magnetic-field-dependent photodynamics of single nv defects in diamond: an application to qualitative all-optical magnetic imaging.
\newblock {\it New Journal of Physics\/} {\bf 14}, 103033 (2012).

\bibitem{Patel2024SinglePhoto-excitation}
A.~Patel, Z.~Chowdhry, A.~Prabhakar, A.~Rathi, V.~P. Bhallamudi, {Single and double quantum transitions in spin-mixed states under photo-excitation}.
\newblock {\it Scientific Reports\/} {\bf 14}, 22421 (2024).

\bibitem{Chen2022DiamagneticResonators}
X.~Chen, S.~K. Ammu, K.~Masania, P.~G. Steeneken, F.~Alijani, {Diamagnetic Composites for High‐Q Levitating Resonators}.
\newblock {\it Advanced Science\/} {\bf 9}, 2203619 (2022).

\bibitem{Tian2024FeedbackPlate}
S.~Tian, K.~Jadeja, D.~Kim, A.~Hodges, G.~C. Hermosa, C.~Cusicanqui, R.~Lecamwasam, J.~E. Downes, J.~Twamley, {Feedback cooling of an insulating high-Q diamagnetically levitated plate}.
\newblock {\it Applied Physics Letters\/} {\bf 124}, 124002 (2024).

\bibitem{Yang2022NVOrientation}
Y.~Yang, H.~H. Vallabhapurapu, V.~K. Sewani, M.~Isarov, H.~R. Firgau, C.~Adambukulam, B.~C. Johnson, J.~J. Pla, A.~Laucht, Observing hyperfine interactions of nv$^{-}$ centers in diamond in an advanced quantum teaching lab.
\newblock {\it American Journal of Physics\/} {\bf 90}, 550--560 (2022).

\bibitem{Romagnoli2023ControllingPlate}
P.~Romagnoli, R.~Lecamwasam, S.~Tian, J.~E. Downes, J.~Twamley, {Controlling the motional quality factor of a diamagnetically levitated graphite plate}.
\newblock {\it Applied Physics Letters\/} {\bf 122}, 094102 (2023).

\end{thebibliography}


\paragraph*{Acknowledgments:}
We acknowledge Pavel Puchenkov with OIST's Scientific Computing and Data Analysis section for assistance with graphics rendering. We thank Prof. K. Dani for the loan of a precision optical chopper. We thank Dr. K. Jadeja and Dr. J. Du for their assistance in initial experiments. We thank Dr. R. Blinder for their insightful discussion on NV polarisation.

\paragraph*{Funding:} The authors acknowledge the Okinawa Institute of Science and Technology Graduate University for funding of this research. S.T. was also supported by JST ASPIRE Grant No. JPMJAP2320.

\end{document}